\documentclass[12pt]{iopart}
\font\cal=cmsy10 scaled 1200
       \newcommand{\Cc}{\mbox{\cal\symbol{'103}}}
       \newcommand{\Dc}{\mbox{\cal\symbol{'104}}}
       \newcommand{\Ec}{\mbox{\cal\symbol{'105}}}
       
       \newcommand{\Hc}{\mbox{\cal\symbol{'110}}}

       \newcommand{\Pc}{\mbox{\cal\symbol{'120}}}
       \newcommand{\Rc}{\mbox{\cal\symbol{'122}}}
\font\semibold=msbm10 scaled 1200

       \newcommand{\R}{\mbox{\semibold\symbol{'122}}}
\font\bfgreek=cmmib10 scaled 1200
       \newcommand{\bftheta}{\mbox{\bfgreek\symbol{'022}}}

\newcommand{\w}{\wedge}

\begin{document}

\title{Contact Equivalence Problem for Nonlinear Wave Equations}

\author{Oleg I. Morozov}

\address{Department of Mathematics, Snezhinsk Physical and Technical Academy,

\noindent
Snezhinsk, 456776, Russia

\noindent
morozov{\symbol{64}}sfti.snz.ru}

\begin{abstract}
The moving coframe method is applied to solve the local equivalence problem
for the class of nonlinear wave equations in two independent variables
under an action of the pseudo-group of contact transformations. The structure
equations and the complete sets of differential invariants for symmetry groups
are found. The solution of the equivalence problem is given in terms of these
invariants.
\end{abstract}

\ams{58H05, 58J70, 35A30}

\section*{Introduction}
In this article we consider a local equivalence problem for the class
of nonlinear second order wave equations
\begin{equation}
w_{tt} = f(x,w_x)\,w_{xx}+g(x,w_x)
\label{IKh1}
\end{equation}
\noindent under a contact transformation pseudo-group. Two
equations are said to be equi\-va\-lent if there exists a contact
transformation mapping one equation to the other. \'Elie Cartan
developed a general method for solving equivalence problems for
submanifolds under an action of a Lie pseudo-group, \cite{Cartan1}
- \cite{Cartan5}. The method provides an effective means of
computing complete systems of differential invariants and
associated invariant differential operators. The necessary and
sufficient condition for equivalence of two sub\-ma\-ni\-folds is
formulated in terms of the dif\-fe\-ren\-ti\-al invariants. The
invariants parameterize the classifying manifold associated with
given submanifolds. Cartan's solution to the equivalence problem
states that two sub\-ma\-ni\-folds are (locally) equivalent if and
only if their classifying manifolds (locally) over\-lap. The
sym\-met\-ry clas\-si\-fi\-ca\-tion problem for classes of
differential equations is closely related to the problem of local
equivalence: symmetry groups and their Lie al\-ge\-bras of two
equations are necessarily isomorphic if these equations are
equivalent, while the converse statement is not true in general.
The preliminary symmetry group classification for the class
(\ref{IKh1}) is given in \cite{ITV}. In \cite{IbragimovKhabirov},
it was proposed to transform equation (\ref{IKh1}) to the
equivalent quasi-linear system of the first order
\begin{equation}
u_t = a (x,u)\, v_x,\qquad v_t = b(x,u)\,u_x,
\label{IKh2}
\end{equation}
\noindent
and the symmetry classification for non-linearizable cases of this system is
given. In \cite{ReidWittkopf} several cases of infinite symmetry algebras
for equation (\ref{IKh1}) are found, and one linearizable case is given.

In the present paper, we apply Cartan's equivalence method,
\cite{Cartan1} - \cite{Cartan5}, \cite{Gardner}, \cite{Olver95},
in its form developed by Fels and Olver, \cite{FO,FO2}, to find
all differential invariants of sym\-met\-ry groups and to solve
the local contact equivalence problem for equations from the class
(\ref{IKh2}) in terms of their coefficients. Unlike Lie's
infinitesimal method, Cartan's approach allows us to find
differential invariants and invariant differential operators
without analyzing over-determined systems of PDEs at all, and
requires differentiation and linear algebra operations only.

The paper is organized as follows. In Section 1, we begin with some notation,
and briefly describe the approach to computing symmetry groups of differential
equations via the moving coframe method of \cite{FO}. In Section 2, the
method is applied to the class of nonlinear wave equations (\ref{IKh2}).
Finally, we make some concluding remarks.

\section{Pseudo-group of contact transformations and symmetries of
differential equations}

In this paper, all considerations are of local nature, and all
mappings are real analytic. Suppose $\Ec = \R^n \times \R^m
\rightarrow \R^n$ is a trivial bundle with the local base
coordinates $(x^1,...,x^n)$ and the local fibre coordinates
$(u^1,...,u^m)$; then by $J^1(\Ec)$ denote the bundle of the
first-order jets of sections of $\Ec$, with the local coordinates
$(x^i,u^{\alpha},p^{\alpha}_i)$, $i\in\{1,...,n\}$,
$\alpha\in\{1,...,m\}$. For every local section
$(x^i,f^\alpha(x))$ of $\Ec$, the corresponding 1-jet
$(x^i,f^\alpha(x),\partial f^\alpha(x)/\partial x^i)$ is denoted
by $j_1(f)$. A differential 1-form $\vartheta$ on $J^1(\Ec)$ is
called a {\it contact form}, if it is annihilated by all 1-jets of
local sections: $j_1(f)^{*}\vartheta = 0$. In the local
coordinates every contact 1-form is a linear combination of the
forms $\vartheta^\alpha = du^\alpha - p^\alpha_{i}\,dx^i$, $\alpha
\in \{1,...,m\}$ (here and later we use the Einstein summation
convention, so $p^\alpha_i\,dx^i =
\sum_{i=1}^{n}\,p^\alpha_i\,dx^i$, etc.) A local diffeomorphism
\begin{equation}
\Delta : J^1(\Ec) \rightarrow J^1(\Ec),
\qquad
\Delta : (x^i,u^\alpha,p^\alpha_i) \mapsto
(\overline{x}^i,\overline{u}^\alpha,\overline{p}^\alpha_i),
\label{Delta}
\end{equation}
\noindent
is called a {\it contact transformation}, if for every contact 1-form
$\vartheta$, the form $\Delta^{*}\overline{\vartheta}$ is also contact,
in other words, if
$\Delta^{*} {\overline \vartheta}^\alpha =
d {\overline u}^\alpha - {\overline p}^\alpha_i \, d {\overline x}^i =
\zeta^\alpha_\beta(x,u,p)\,\vartheta^\beta$ for some functions
$\zeta^\alpha_\beta$ on $J^1(\Ec)$.

Cartan's method of equivalence, \cite{Cartan2,Cartan5,Olver95}, allows us to
compute invariant 1-forms which define the pseudo-group of contact
transformations. The result of its application is the following (see
\cite{Morozov}). Consider the lifted coframe
\begin{eqnarray}
\Theta^\alpha = a^\alpha_\beta\, (d u^\beta - p^\beta_j\,d x^j),\nonumber\\
\Xi^i = c^i_\beta\, \Theta^\beta +b^i_j\, d x^j,  \label{LCF}\\
\Sigma^\alpha_i = f^\alpha_{i\beta}\,\Theta^\beta+g^\alpha_{ij}\, \Xi^j +
a^\alpha_\beta\,B^j_i\,\,d p^\beta_j \nonumber
\end{eqnarray}
\noindent
on $J^1(\Ec) \times \Hc$, where $\Hc$ is the Lie group of block lower
triangular matrices of the form
\[
\left(
\begin{array}{ccc}
a^\alpha_\beta & 0 & 0\\
c^i_\gamma\,a^\gamma_\beta & b^i_j & 0\\
(f^\alpha_{i\gamma}+g^\alpha_{ik}\,c^k_\gamma)\,a^\gamma_\beta&
g^\alpha_{ik}\,b^k_j&a^\alpha_\beta\,B^j_i
\end{array}
\right),
\]
\noindent
and the parameters $a^\alpha_\beta$, $b^i_j$, $c^i_\beta$, $f^\alpha_{i\beta}$,
and $g^\alpha_{ij}$ obey the requirements
${\mbox{\rm{det}}} \,\left(a^\alpha_\beta\right)\not = 0$,
${\mbox{\rm{det}}} \,\left(b^i_j\right)\not = 0$,
$b^i_k\,B^k_j = \delta^i_j$, and $g^\alpha_{ij}=g^\alpha_{ji}$.
Then a transformation
$\Upsilon : J^1(\Ec)\times \Hc  \rightarrow J^1(\Ec)\times \Hc$
satisfies the conditions
\[
\Upsilon^{*}\,\overline \Theta^\alpha=\Theta^\alpha, \qquad
\Upsilon^{*}\,\overline \Xi^i=\Xi^i, \qquad
\Upsilon^{*}\,\overline \Sigma^\alpha_i=\Sigma^\alpha_i
\]
\noindent if and only if it is projectable on $J^1(\Ec)$
and its projection
$\Delta : J^1(\Ec) \rightarrow J^1(\Ec)$
is a contact transformation.
The lifted coframe has the structure equations
\begin{eqnarray}
d \Theta^\alpha = \Phi^\alpha_\beta \wedge \Theta^\beta
+ \Xi^k \wedge \Sigma^\alpha_k,
\nonumber\\
d \Xi^i = \Psi^i_k\wedge \Xi^k +\Pi^i_\gamma \wedge \Theta^\gamma,
\label{StructureEquations}\\
d \Sigma^\alpha_i = \Phi^\alpha_\gamma\wedge\Sigma^\gamma_i
-\Psi^k_i\wedge\Sigma^\alpha_k+\Lambda^\alpha_{i\beta}\wedge\Theta^\beta
+\Omega^\alpha_{ij}\wedge\Xi^j,\nonumber
\end{eqnarray}
\noindent
where $\Phi^\alpha_\beta$, $\Psi^i_k$, $\Pi^i_\gamma$,
$\Lambda^\alpha_{i\beta}$, and $\Omega^\alpha_{ij}$ are 1-forms on
$J^1(\Ec) \times \Hc$, and, as it is shown in \cite{Morozov}, the coframe is
involutive.

The structure equations (\ref{StructureEquations}) remain unchanged
if we make the following change of the modified Maurer - Cartan forms
$\Phi^\alpha_\beta$, $\Psi^i_k$, $\Pi^i_\gamma$, $\Lambda^\alpha_{i\beta}$,
and $\Omega^\alpha_{ij}$~ :
\begin{equation}
\begin{array}{lll}
\Phi^\alpha_\beta &\mapsto& \Phi^\alpha_\beta
+ K^\alpha_{\beta\gamma}\,\Theta^\gamma,
\\
\Psi^i_k &\mapsto& \Psi^i_k + L^i_{kj}\,\Xi^j+M^i_{k\gamma}\,\Theta^\gamma,
\\
\Pi^i_\gamma &\mapsto& \Pi^i_\gamma + M^i_{k\gamma}\,\Xi^k
+N^i_{\gamma \epsilon}\,\Theta^\epsilon,
\\
\Lambda^\alpha_{i\beta} &\mapsto& \Lambda^\alpha_{i\beta} +
P^\alpha_{i\beta\gamma}\,\Theta^\gamma+Q^\alpha_{i\beta k}\,\Xi^k
+ K^\alpha_{\gamma\beta}\,\Sigma^\gamma_i-M^k_{i\beta}\,\Sigma^\alpha_k,
\\
\Omega^\alpha_{ij} &\mapsto& \Omega^\alpha_{ij}
+ Q^\alpha_{i\beta j}\,\Theta^\beta
+R^\alpha_{ijk}\,\Xi^k - L^k_{ij}\,\Sigma^\alpha_k,
\end{array}
\label{Gamma_rules}
\end{equation}
\noindent
where $K^\alpha_{\gamma\epsilon}$, $L^i_{kj}$, $M^i_{k\gamma}$,
$N^i_{\gamma \epsilon}$, $P^\alpha_{i\beta\gamma}$, $Q^\alpha_{i\beta k}$,
and $R^\alpha_{ijk}$ are arbitrary functions on
$J^1(\Ec)\times \Hc$  satisfying the following symmetry
conditions:
\begin{eqnarray*}
K^\alpha_{\gamma\epsilon}=K^\alpha_{\epsilon\gamma},~~
L^i_{kj}=L^i_{jk},~~
N^i_{\gamma \epsilon}=N^i_{\epsilon\gamma}, \\
P^\alpha_{i\beta\gamma}=P^\alpha_{i\gamma\beta},~~
Q^\alpha_{i\beta k}=Q^\alpha_{k\beta i},~~
R^\alpha_{ijk}=R^\alpha_{ikj}=R^\alpha_{jik}.
\end{eqnarray*}

Another approach to construct 1-forms characterizing contact transformations
is presented in \cite{Olver2000}.

Suppose $\Rc$ is a first-order differential equation in $m$ dependent and
$n$ independent variables. We consider $\Rc$ as a sub-bundle in $J^1(\Ec)$.
Let $Cont(\Rc)$ be the group of contact symmetries for $\Rc$. It consists of
all the contact transformations on $J^1(\Ec)$ mapping $\Rc$ to itself.
The moving coframe method, \cite{FO,FO2}, is applicable to find invariant
1-forms characterizing $Cont(\Rc)$ is the same way, as the lifted coframe
(\ref{LCF}) to $J^1(\Ec)\times\Hc$ characterizes $Cont(J^1(\Ec))$. We briefly
outline this approach.

Let $\iota : \Rc \rightarrow J^1(\Ec)$ be an embedding. The invariant
1-forms of $Cont(\Rc)$ are re\-stric\-ti\-ons of the coframe (\ref{LCF}) on
$\Rc$: $\theta^\alpha = \iota^{*} \Theta^\alpha$, $\xi^i = \iota^{*}\Xi^i$,
and $\sigma^\alpha_i=\iota^{*}\Sigma^\alpha_i$ (for brevity we identify the
map $\iota \times id : \Rc\times \Hc \rightarrow J^1(\Ec)\times \Hc$ with
$\iota : \Rc \rightarrow J^1(\Ec)$). The forms $\theta^\alpha$, $\xi^i$, and
$\sigma^\alpha_i$ have some linear dependencies, i.e., there exists a
non-trivial set of functions $S_{\alpha}$, $T_{i}$, and $U_{\alpha}^i$ on
$\Rc\times \Hc$ such that
$S_{\alpha}\,\theta^\alpha+T_{i}\,\xi^i+U_{\alpha}^i\,\sigma^\alpha_i\equiv 0$.
These functions are lifted invariants of $Cont(\Rc)$. Setting them equal to
appropriate constants allows us to specify some parameters $a^\alpha_\beta$,
$b^i_j$, $c^i_\beta$, $f^\alpha_{i\beta}$, and $g^\alpha_{ij}$ of the group
$\Hc$ as functions of the coordinates on $\Rc$ and the other group parameters.

After these normalizations, some restrictions of the forms
$\phi^\alpha_\beta=\iota^{*}\Phi^\alpha_\beta$,
$\psi^i_k=\iota^{*}\Psi^i_k$, $\pi^i_\beta=\iota^{*}\Pi^i_\beta$,
$\lambda^\alpha_{i\beta}=\iota^{*}\Lambda^\alpha_{i\beta}$, and
$\omega^\alpha_{ij}=\iota^{*}\Omega^\alpha_{ij}$, or some their
linear combinations, become semi-basic, i.e., they do not include
the differentials of the parameters of $\Hc$. From
(\ref{Gamma_rules}), we have the following statements: (i) if
$\phi^\alpha_\beta$ is semi-basic, then its coefficients at
$\sigma^\gamma_j$ and $\xi^j$ are lifted invariants of
$Cont(\Rc)$; (ii) if $\psi^i_k$ or $\pi^i_\beta$ are semi-basic,
then their coefficients at $\sigma^\gamma_j$ are lifted invariants
of $Cont(\Rc)$. Setting these invariants equal to some constants,
we get specifications of some more parameters of $\Hc$ as
functions of the coordinates on $\Rc$ and the other group
parameters.

More lifted invariants can appear as essential torsion coefficients in the
reduced structure equations
\begin{eqnarray*}
d \theta^\alpha = \phi^\alpha_\beta \wedge \theta^\beta
+ \xi^k \wedge \sigma^\alpha_k,
\\
d \xi^i = \psi^i_k\wedge \xi^k
+\pi^i_\gamma \wedge \theta^\gamma,
\\
d \sigma^\alpha_i = \phi^\alpha_\gamma\wedge\sigma^\gamma_i
-\psi^k_i\wedge\sigma^\alpha_k
+\lambda^\alpha_{i\beta}\wedge\theta^\beta
+\omega^\alpha_{ij}\wedge\xi^j.
\end{eqnarray*}
\noindent
After normalizing these invariants and repeating the process, two outputs are
possible. In the first case, the reduced lifted coframe appears to be
involutive. Then this coframe is the desired set of defining forms for
$Cont(\Rc)$. In the second case, when the reduced lifted coframe does not
satisfy Cartan's test, we should use the procedure of prolongation,
\cite[ch~12]{Olver95}.

\section{Structure and invariants of symmetry groups for nonlinear wave
equations}

We apply the method described in the previous section to the class of nonlinear
wave equations (\ref{IKh2}). Denote $x^1=t$, $x^2=x$, $u^1=u$, $u^2=v$,
$p^1_1=u_t$,
$p^1_2=u_x$,
$p^2_1=v_t$,
and
$p^2_2=v_x$,
The coordinates on $\Rc$ are $\{t, x, u, v, u_x, v_x\}$,
and the embedding $\iota : \Rc \rightarrow J^1(\Ec)$ is defined by
(\ref{IKh2}). For simplicity in the following computations, we put
$F(x,u) = (a(x,u)\,b(x,u))^{1/2}$ and
$G(x,u) = (b(x,u)/a(x,u))^{1/2}$,
so $a(x,u)=F(x,u)/G(x,u)$ and
$b(x,u)=F(x,u)\,G(x,u)$.

There are three cases to be treated separately: Case A, when
$F_u\not =0$ and $G_x \not = 0$, Case B, when $G_x = 0$, and Case
C, when $F_u = 0$.

In the case B system (\ref{IKh2}) has the form
$u_t = F(x,u)\,(G(u))^{-1}\,v_x$, $v_t = F(x,u)\,G(u)\,u_x$,
so the change of variables $\tilde{u}=H(u)$ provided $H^{\prime}(u)=G(u)$
transforms this system into the system
$\tilde{u}_t = \tilde{F}(x,\tilde{u})\,v_x$,
$v_t = \tilde{F}(x,\tilde{u})\,\tilde{u}_x$
with $\tilde{F}(x,\tilde{u})=F(x,H^{-1}(\tilde{u}))=F(x,u)$.
Therefore we drop the tildes and conclude that in the case B system
(\ref{IKh2}) is equivalent to the system
\begin{equation}
u_t = F(x,u)\,v_x,   \qquad  v_t = F(x,u)\,u_x.
\label{B}
\end{equation}
\noindent

In the case C system (\ref{IKh2}) has the form
$u_t = F(x)\,(G(x,u))^{-1}\,v_x$, $v_t = F(x)\,G(x,u)\,u_x$,
so the change of variables $\tilde{x}=H(x)$ provided $H^{\prime}(x)=1/F(x)$
transforms this system into the system
$u_t = (\tilde{G}(\tilde{x},u))^{-1}\,v_{\tilde{x}}$,
$v_t = \tilde{G}(\tilde{x},u)\,u_{\tilde{x}}$,
with $\tilde{G}(\tilde{x},u)=G(H^{-1}(\tilde{x}),u)=G(x,u)$.
Next, the contact transformation $\overline{t} = v$, $\overline{x} = u$,
$\overline{u} = \tilde{x}$, and $\overline{v} = t$ maps the last system to the
system in the form
$\overline{u}_{\overline{t}} =
\overline{F}(\overline{x},\overline{u})\,\overline{v}_{\overline{x}}$,
$\overline{v}_{\overline{t}} =
\overline{F}(\overline{x},\overline{u})\,\overline{u}_{\overline{x}}$.
Thus in the case C system (\ref{IKh2}) is equivalent under a contact
transformation to the system in the form (\ref{B}) too.

Let us analyze the system
\begin{equation}
u_t = F(x,u)\,(G(x,u))^{-1}\,v_x,\qquad
v_t = F(x,u)\,G(x,u)\,u_x
\label{Main}
\end{equation}
\noindent
in the case A. For brevity we denote
\begin{equation}
P = {{G_x\,F^2}\over{F_u}}.
\label{P}
\end{equation}
Computing the linear dependence conditions for the reduced forms
$\theta^\alpha$, $\xi^i$, and $\sigma^\alpha_i$ by means
of {\sc MAPLE}, we express the group parameters $a^1_2$, $a^2_1$,
$b^1_2$, $b^2_1$, $f^1_{11}$, $f^1_{12}$, $f^2_{21}$, $f^2_{22}$, $g^1_{11}$,
$g^1_{12}$, $g^2_{12}$, and $g^2_{22}$. Particularly, since
\[
\sigma^1_1\equiv
{{
F\,(a^1_1-G\,a^1_2)\,(a^1_1+G\,a^1_2)\,\det(a^\alpha_\beta)
}\over{
G\,(b^1_1-G\,b^1_2)\,(b^1_1+G\,b^1_2)\,\det(b^i_j)
}}
\,\sigma^2_2
\quad({\mbox{\rm{mod}}}\,\theta^1, \theta^2, \xi^1,\xi^2, \sigma^1_2),
\]
\noindent
we take $a^1_2 = G^{-1}\,a^1_1$. Then
\[
\sigma^1_1\equiv
{{
(b^2_1-F\,b^2_2)
}\over{
(b^1_2-F\,b^1_1)
}}
\,\sigma^1_2
\quad({\mbox{\rm{mod}}}\,\theta^1, \theta^2, \xi^1,\xi^2),
\]
\noindent
and we take $b^2_1 =F\,b^2_2$. Similarly, we set the coefficients of
$\sigma^1_1$ at $\theta^1$, $\theta^2$, $\xi^1$, and $\xi^2$ equal to zero,
and express $f^1_{11}$, $f^1_{12}$, $g^1_{11}$, and $g^1_{12}$, respectively.

Then we obtain
\[
\sigma^2_1\equiv
{{
F\,(a^2_1+G\,a^2_2)\,b^2_2
}\over{
(b^1_2+F\,b^1_1)\,a^1_1
}}
\,\sigma^1_2
\quad({\mbox{\rm{mod}}}\,\theta^1, \theta^2, \xi^1,\xi^2, \sigma^2_2),
\]
\noindent
so we take $a^1_2 = - G\,a^2_2$. Now we get
\[
\sigma^2_1\equiv
-{{
2\,F\,b^2_2
}\over{
(b^1_1+F\,b^1_2)
}}
\,\sigma^2_2
\quad({\mbox{\rm{mod}}}\,\theta^1, \theta^2, \xi^1,\xi^2).
\]
\noindent
Since $b^2_2 \not = 0$ (otherwise $b^2_1= 0$ and $\det(b^i_j) = 0$), we
set the coefficient at $\sigma^2_2$ equal to 1, and obtain
$b^1_2 = -(F^{-1}\,b^1_1+2\,b^2_2)$.  After that, we set the coefficients of
$\sigma^2_1$ at $\theta^1$, $\theta^2$, $\xi^1$, and $\xi^2$ equal to zero
and find $f^2_{21}$, $f^2_{22}$, $g^2_{12}$, and $g^2_{22}$, respectively.
This yields
\begin{equation}
\sigma^1_1=0,\qquad \sigma^2_1 = \sigma^2_2.
\label{lin_dependence}
\end{equation}

At the next step we analyze the forms $\phi^\alpha_\beta
=\iota^{*}\,\Phi^\alpha_\beta$ and $\psi^i_j =\iota^{*}\,\Psi^i_j$
reduced by setting (\ref{lin_dependence}) and substituting the
values of $a^1_2$, $a^2_1$, $b^1_2$, $b^2_1$, $f^1_{11}$,
$f^1_{12}$, $f^2_{21}$, $f^2_{22}$, $g^1_{11}$, $g^1_{12}$,
$g^2_{12}$, and $g^2_{22}$ obtained at the previous step. The form
$\phi^1_2$ is semi-basic now, and $\phi^1_2 \equiv
c^2_2\,\sigma^1_2 ({\rm{mod}}\,\theta^1, \theta^2, \xi^1, \xi^2,
\sigma^2_2)$. So we take $c^2_2 = 0$. For the semi-basic form
$\phi^2_1$ we have $\phi^2_1 \equiv (c^2_1+c^1_1)\,\sigma^1_2
({\rm{mod}}\,\theta^1, \theta^2, \xi^1, \xi^2, \sigma^2_2)$, so we
put $c^2_1 = - c^1_1$. Then
\[
\phi^1_2 \equiv
{{
F_u\,(P-F\,G\,u_x-F\,v_x)\,a^1_1
}\over{
4\,F\,G^2\,a^2_2\,(b^1_1+F\,b^2_2)
}}\,\xi^1
\quad
({\mbox{\rm{mod}}}\,\theta^1, \theta^2, \xi^2),
\]
\noindent
and we take
$a^2_2 =
F_u\,a^1_1\,(P-F\,G\,u_x-F\,v_x)\,F^{-1}\,G^{-2}\,(b^1_1+F\,b^2_2)^{-1}$.
Then we have the semi-basic linear combination
$\psi^1_1-\phi^1_1+\phi^2_2$ with
\[\fl\hspace{20pt}
\psi^1_1-\phi^1_1+\phi^2_2\equiv
{{
(P-F\,G\,u_x-F\,v_x)\,a^1_1\,c^1_1-F\,G\,b^2_2
}\over{
(P-F\,G\,u_x-F\,v_x)\,a^1_1
}}\,\sigma^1_2
\quad
({\mbox{\rm{mod}}}\,\theta^1, \theta^2, \xi^1, \xi^2),
\]
\noindent
so we take $c^1_1 = F\,G\,b^2_2\,(P-F\,G\,u_x-F\,v_x)^{-1}\,(a^1_1)^{-1}$.
Similarly, we set the coefficients of $\phi^1_2$ and $\phi^2_1$ at $\xi^2$
equal to zero, and find $f^1_{22}$ and $f^2_{11}$, respectively. Then
\[
\phi^2_1 \equiv
{{
F_u^2\,(P-F\,G\,u_x-F\,v_x)\,(P-F\,G\,u_x+F\,v_x)
}\over{
4\,F^3\,G^2\,b^2_2\,(b^1_1+F\,b^2_2)
}}\,
\xi^1
\quad
({\mbox{\rm{mod}}}\,\theta^1, \theta^2),
\]
\noindent
so we set the coefficient at $\xi^1$ equal to 1 and find
\[
b^1_1 =
{{
F_u^2\,(P-F\,G\,u_x-F\,v_x)\,(P-F\,G\,u_x+F\,v_x)-4\,F^4\,G^2\,(b^2_2)^2
}\over{
4\,F^3\,G^2\,b^2_2
}}.
\]
Then the semi-basic linear combination $\psi^1_2 +2\,(\phi^2_2-\phi^1_1)$
gives
\[\fl
\psi^1_2+2(\phi^2_2-\phi^1_1)\equiv
{{
16F^5G^2(b^2_2)^2c^1_2 - F_u^3
((P-F\,G\,u_x)^2-F^2\,v_x^2)
}\over{
16\,F^5\,G^2\,a^1_1\,(b^2_2)^2
}}\,
\sigma^2_2
\quad
({\mbox{\rm{mod}}}\,\theta^1, \theta^2, \xi^1, \xi^2),
\]
therefore we put
\[
c^1_2 =
-{{
F_u^3\,((P-F\,G\,u_x)^2-F\,v_x^2)
}\over{
16\,F^5\,G^2\,a^1_1\,(b^2_2)^2
}}.
\]

At the third step, we analyze the reduced structure equations.
After absorption, we have an essential torsion coefficient at
$\xi^1\w\sigma^1_2$ in $d\sigma^1_2$. This coefficient depends on
$f^2_{12}$; we set the coefficient equal to zero and express
$f^2_{12}$, while the expression is too long to be written in
full. Similarly, we express $f^2_{21}$ from the essential torsion
coefficient at $\xi^2\w\sigma^2_2$ in $d\sigma^2_2$. Then after
absorption of torsion in all the structure equations we have
\[\fl
d\sigma^2_2 =
\zeta_1\w(2\,\theta^1+\sigma^2_2)+\zeta_2\w(\theta^1+\sigma^2_2)
+\zeta_3\w\theta^2+\zeta_4\w(\xi^1+\xi^2)-\xi^2\w\sigma^2_2
\]
\[\fl\hspace{20pt}
-{{
2\,F^5\,G^2\,(b^2_2)^2\,\left(G\,P_x+(G\,u_x-v_x)\,P_u\right)
}\over{
F_u^2\,a^1_1\,(P-FGu_x+Fv_x)\,(P-FGu_x-Fv_x)^3
}}\,
\theta^2\w\sigma^2_2,
\]
where $\zeta_1$, $\zeta_2$, $\zeta_3$, and $\zeta_4$ are 1-forms on
$\Rc\times\Hc$, and the last torsion coefficient is essential. There are two
possibilities now: $P \not = const$ and $P = const$. Denote by $\Pc_1$ the
subclass of systems (\ref{Main}) such that $G_x\not = 0$, $F_u\not = 0$, and
$P \not = const$. For a system from $\Pc_1$ we set the coefficient at
$\theta^2\w\sigma^2_2$ in $d\sigma^2_2$ equal to 1 and obtain
\[\fl\hspace{20pt}
a^1_1 = -{{
2\,F^5\,G^2\,(b^2_2)^2\,\left(G\,P_x+(G\,u_x-v_x)\,P_u\right)
}\over{
F_u^2\,(P-FGu_x+Fv_x)\,(P-FGu_x-Fv_x)^3
}}.
\]
\noindent Similarly, we set the essential torsion coefficient at
$\theta^1\w\theta^2$ in $d\xi^1$ equal to zero and find
\[
b^2_2=
-{{
F_u\,(P-FGu_x-Fv_x)
}\over{
2\,F^2G
}}.
\]
\noindent
Next, we express $g^1_{22}$ from the essential torsion coefficient
at $\theta^1\w\xi^1$ in $d\sigma^1_2$. Now the essential torsion coefficient
at $\theta^1\w\sigma^1_2$ in $d\theta^1$ has the form
\[
R=
{{
F_u\,(FGP_x+PP_u)\,(P-FGu_x+Fv_x)^2
}\over{
2\,F^3\,\left(G\,P_x+(G\,u_x-v_x)\,P_u\right)^2
}}.
\]
\noindent This function is an invariant of the symmetry group for
a system from $\Pc_1$, together with its invariant derivatives
$\Dc_i(R)$, $i\in\{1,...,6\}$, defined by $d R
=\Dc_1(R)\,\theta^1+\Dc_2(R)\,\theta^2+\Dc_3(R)\,\xi^1
+\Dc_4(R)\,\xi^2+\Dc_5(R)\,\sigma^1_2+\Dc_6(R)\,\sigma^2_2$. The
invariant $\Dc_3(R)$ depends on $g^2_{11}$; we set $\Dc_3(R)=0$
and express $g^2_{11}$.

Now all the parameters of the group $\Hc$ are expressed as functions of
$x$, $u$, $u_x$, and $v_x$. The structure equations of the symmetry
group for a system from  $\Pc_1$ have the form

\[\fl
d\theta^1=
\case16\,(6\,K_3+1-4\,K_2K_3K_5-2\,K_4)\,
\theta^1\w\theta^2
\]
\[\fl\hspace{20pt}
+\case16\,(K_4+1+2\,K_2K_3K_5-6\,K_3)\,K_2^{-1}K_3^{-1}\,
\theta^1\w\xi^1
\]
\[\fl\hspace{20pt}
+\case13\,(K_4+1+3\,K_1K_2K_3+3\,K_2K_3K_6-6\,K_3-4\,K_2K_3K_5)\,
K_2^{-1}K_3^{-1}\,\theta^1\w\xi^2
\]
\[\fl\hspace{20pt}
+K_2K_3\,\theta^1\w\sigma^2_2
-\case14\,\theta^2\w\xi^1
+\xi^2\w\sigma^1_2,
\]
\[\fl
d\theta^2=
K_4\,\theta^1\w\theta^2
-\theta^1\w\xi^1
+K_5\,\theta^2\w\xi^1
+K_6\,\theta^2\w\xi^2
+K_2K_3\,\theta^2\w\sigma^2_2
+(\xi^1+\xi^2)\w\sigma^2_2,
\]
\[\fl
d\xi^1=
K_1K_2\,\theta^1\w\xi^1
+\case12\,K_1K_2\,\theta^2\w\xi^1
+K_1\,\xi^1\w\xi^2,
\]
\[\fl
d\xi^2=
K_2K_3\,\theta^1\w\theta^2
-\theta^1\w\xi^1
+(K_1K_2-1)\,\theta^1\w\xi^2
+\case12\,(2\,K_3-1+K_1K_2)\,\theta^2\w\xi^2
\]
\[\fl\hspace{20pt}
+\case16\,(K_4+1-4\,K_2K_3K_5-6\,K_3)\,K_2^{-1}K_3^{-1}\,
\xi^1\w\xi^2,
\]
\[\fl
d\sigma^1_2=
-K_{10}\,\theta^1\w\theta^2
-K_{11}\,\theta^1\w\xi^2
+(K_4-1)\,\theta^1\w\sigma^1_2
\]
\[\fl\hspace{20pt}
+\case1{12}\,
(6\,K_3-1+4\,K_2K_3K_5-K_4)\,
K_2^{-1}K_3^{-1}\,
\theta^2\w\xi^1
\]
\[\fl\hspace{20pt}
-K_9\,\theta^2\w\xi^2
+\case16\,(2\,K_4-3\,K_1K_2+4\,K_2K_3K_5-12\,K_3+2)\,
\theta^2\w\sigma^1_2
\]
\[\fl\hspace{20pt}
+K_8\,\xi^1\w\xi^2
-\case13\,(K_4-6\,K_3+1-K_2K_3K_5)\,K_2^{-1}K_3^{-1}\,
\xi^1\w\sigma^1_2
+\case14\,\xi^1\w\sigma^2_2
\]
\[\fl\hspace{20pt}
+K_7\,\xi^2\w\sigma^1_2
-K_{12}\,\xi^2\w\sigma^2_2
+K_2K_3\,\sigma^1_2\w\sigma^2_2,
\]
\[\fl
d\sigma^2_2=
K_{19}\,\theta^1\w\theta^2
+K_{17}\w\theta^1\w\xi^1
+K_{14}\,\theta^1\w\xi^2
+(K_4-K_1K_2+1)\,\theta^1\w\sigma^2_2
\]
\[\fl\hspace{20pt}
+K_{18}\,\theta^2\w\xi^1
+K_{15}\,\theta^2\w\xi^2
+\case13\,(2\,K_2K_3K_5+K_4-2)\,\theta^2\w\sigma^2_2
-K_{16}\,\xi^1\w\xi^2
\]
\[\fl\hspace{20pt}
+K_{13}\,\xi^1\w\sigma^2_2
-\xi^2\w\sigma^1_2
-K_{20}\,\xi^2\w\sigma^2_2,
\]
\noindent
where
\[\fl
K_1={\frac {2Fv_{x}\left (2\,F_{u}^2GP+F_uG_uFP-F_{uu}F G P
-G^2F^2F_{xu}-F_uFGP_u+G^2F_xF_uF\right)} {\left
(P-u_xGF+Fv_x\right )F_u^2G\left (-P+u_xGF+Fv_x\right )}},
\]
\[\fl
K_2={\frac {F_{{u}}\,\left (-P+u_{{x}}GF+Fv_{{x}} \right )\left
(-P+u_{{x}}G F-Fv_{{x}}\right )^{2}}{2{F}^{3}v_{{x}} \left
(u_{{x}}P_uG+P_xG-P_uv_{{x}}\right )}},
\]
\[\fl
K_3=
{{v_x\,(PP_u+FGP_x)}\over{(FGu_x+Fv_x-P)\,(GP_x+GP_uu_x-P_uv_x)}},
\]
\noindent while the expressions for $K_4$, ..., $K_{20}$ are too
long to be written in full.

The functions $K_1$, ..., $K_{20}$ are differential invariants of
the symmetry group $Cont(\Rc)$ for system (\ref{Main}) from
$\Pc_1$. All the other differential invariants of $Cont(\Rc)$ are
functions of $K_j$ and their invariant derivatives $K_{j,I} =
\Dc_I(K_j)$, where for a multi-index $I=(i_1,i_2,...,i_l)$ of
length $\# I = l$ we denote $\Dc_I = \Dc_{i_1}\circ \Dc_{i_2}
\circ ... \circ \Dc_{i_l}$, $i_k \in \{1,...,6\}$ for
$k\in\{1,...,l\}$. For $s\ge0$ the $s^{th}$ order {\it classifying
manifold} associated with the coframe $\bftheta =
\{\theta^1,\theta^2,\xi^1,\xi^2,\sigma^1_2,\sigma^2_2\}$ and an
open subset $V$ in space $\R^4$ with the coordinates
$(x,u,u_x,v_x)$ is
\begin{equation}\fl
\Cc^{(s)}(\bftheta,V) = \{(K_{j,I}(x,u,u_x,v_x))\,\,\vert\,\,
j\in\{1,...,20\},\,\,
\#I\le s, \,\,(x,u,u_x,v_x)\in V\}.
\label{manifold_A}
\end{equation}
\noindent
Since all the functions $K_{j,I}$ depend on four variables $x$, $u$, $u_x$,
and $v_x$, it follows that $\rho_s = \dim \Cc^{(s)}(\bftheta,V) \le 4$ for all
$s\ge 0$. Let $r=\min \{s \,\,\vert\,\, \rho_s=\rho_{s+1}=\rho_{s+2}= ...\}$
be the {\it order of the coframe} $\bftheta$. We have
$0\le \rho_0\le\rho_1\le\rho_2 \le ... \le 4$. In any case, $r+1 \le 4$.
Hence from Theorem 8.19 of \cite{Olver95} we see that two systems (\ref{Main})
from the subclass $\Pc_1$ are locally equivalent under a contact transformation
if and only if their fourth order classifying manifolds (\ref{manifold_A})
locally overlap. The dimension of $Cont(\Rc)$ is equal to
$6 - \dim \Cc^{(4)}(\bftheta,V)$. Therefore $\dim Cont(\Rc) \ge 2$, as it
should be, since every system (\ref{Main}) is invariant under the symmetries
with infinitesimal generators $\partial / \partial t$ and
$\partial / \partial v$.

\vskip 5 pt

Now we consider the case $G_x \not = 0$, $F_u\not = 0$, and $P = m = const$.
From (\ref{P}) it follows that the system $H_x = -m\,F^{-1}$, $H_u = G$ is
compatible, therefore there exists a function $H(x,u)$ such that
$dH = - m\,F^{-1}\,dx + G\,du$. Then the change of variables
$\tilde{u} = H(x,u)$, $\tilde{v} = v-m\,t$ maps system (\ref{Main}) to the
system $\tilde{u}_t = \tilde{F}(x,\tilde{u})\,\tilde{v}_x$,
$\tilde{v}_t = \tilde{F}(x,\tilde{u})\,\tilde{u}_x$
with $\tilde{F}(x,\tilde{u}) = F(x,u)$. Dropping tildes, we obtain
system (\ref{B}). Thus in the case $P=const$ system (\ref{Main})
is equivalent to system (\ref{B}).

Let us consider system (\ref{B}). The computations are similar, so we omit
them and present the results. The structure of the symmetry group for system
(\ref{B}) is different in the cases of $(\ln F)_{xu} \not = 0$ and
$(\ln F)_{xu} = 0$. We denote by $\Pc_2$ the subclass of systems (\ref{B})
such that $(\ln F)_{xu} \not = 0$. For a system from $\Pc_2$ all the parameters
of the group $\Hc$ are functions of $x$, $u$, $u_x$, and $v_x$.
The structure equations for the coframe $\bftheta$ have the form
\[\fl
d\theta^1=
(L_{{3}}\,\theta^1+\xi^1)\w\theta^2
+\case13\, (3\,L_{{2}}L_{{3}}+L_{{2}}-L_{{4}}+L_{{1}}
-L_{{1}}L_{{3}})\,\theta^1\w\xi^1
+(L_{{4}}\,\theta^1-\sigma^1_2)\w\xi^2,
\]
\[\fl
d\theta^2=
\case12\,L_{{3}}\,\theta^1\w\theta^2
+\theta^1\w\xi^2
+\case13\,(3\,L_{{2}}L_{{3}}-2\,L_{{2}}-L_{{4}}+L_{{1}}-L_{{1}}L_{{3}})\,
\theta^2\w\xi^1
\]\[\fl\hspace{20pt}
-(2\,L_{{2}}+L_{{1}}-L_{{4}})\,\theta^2\w\xi^2
+(\xi^1+\xi^2)\w\sigma^2_2,
\]
\[\fl
d\xi^1=
-\case12\,\theta^1\w\xi^1
+\theta^2\w\xi^1
+L_{{1}}\,\xi^1\w\xi^2,
\]
\[\fl
d\xi^2=
-\case12\,\theta^1\w\xi^2
+\theta^2\w\xi^2
+L_{{2}}\,\xi^1\w\xi^2,
\]
\[\fl
d\sigma^1_2=
\case13\,(4\,L_{{2}}+3\,L_{{2}}L_{{3}}-L_{{3}}L_{{1}}-4\,L_{{4}}+6
+4\,L_{{1}})\,(\theta^1\w\theta^2-\xi^2\w\sigma^1_2)
+L_{{7}}\,\theta^1\w\xi^1
\]\[\fl\hspace{20pt}
+(2\,L_{{2}}L_{{3}}-\case{14}3\,L_2^{2}-2\,L_2^{2}L_{{3}}
+\case13\,L_{{2}}L_{{4}}+\case1{13}\,L_{{2}}-\case13\,L_{{1}}L_{{2}}L_{{3}}
-7\,L_{{2}}L_{{1}}-\case{25}{6}\,L_{{4}}+\case13\,{L_{{1}}}^{2}L_{{3}}
\]\[\fl\hspace{20pt}
+\case73\,L_{{1}}L_{{4}}-\case23\,L_{{3}}L_{{1}}+1+\case{25}{6}\,L_{{1}}
-\case12\,L_{{6}}-\case73\,L_1^{2})\,
\theta^1\w\xi^2
+\case12\,L_{{3}}\,\theta^1\w\sigma^1_2
-2\,L_{{2}}\,\theta^2\w\xi^1
\]\[\fl\hspace{20pt}
+\theta^2\w(L_{{6}}\,\xi^2-(L_{{3}}+1)\,\sigma^1_2)
+\case13\,(14\,L_{{2}}+6\,L_{{2}}L_{{3}}-2\,L_{{3}}L_{{1}}-8\,L_{{4}}+3
+14\,L_{{1}})\,\xi^2\w\sigma^2_2
\]\[\fl\hspace{20pt}
+L_{{5}}\,\xi^1\w\xi^2
+\case13\, (L_{{4}}-3\,L_{{2}}L_{{3}}-4\,L_{{2}}-L_{{1}}+L_{{1}}L_{{3}})\,
\xi^1\w\sigma^1_2
+\xi^1\w\sigma^2_2,
\]
\[\fl
d\sigma^2_2=
-\case16\,(3\,L_{{2}}L_{{3}}+L_{{2}}+\case32+L_{{1}}-L_{{1}}L_{{3}}
-L_{{4}})\,\theta^1\w\theta^2
+\case1{12}\,(2\,L_1^{2}-2\,L_1^{2}L_{{3}}+4\,L_{{1}}L_{{3}}
\]\[\fl\hspace{20pt}
+2\,L_{{1}}L_{{2}}L_{{3}}-2\,L_{{1}}L_{{4}}-L_{{1}}
+6\,L_{{1}}L_{{2}}+6\,L_{{8}}+2\,L_{{2}}L_{{4}}+L_{{4}}-6
-12\,L_{{2}}L_{{3}}+4\,L_{{2}}^{2}
\]\[\fl\hspace{20pt}
+12\,L_{{2}}^{2}L_{{3}}+2\,L_{{2}})\,
\theta^1\w(\xi^1-\xi^2)
+\case12\, (L_{{3}}+1)\,\theta^1\w\sigma^2_2
-(L_{{1}}-L_{{8}}+L_{{7}}-3\,L_{{2}})\,\theta^2\w\xi^1
\]\[\fl\hspace{20pt}
+L_{{8}}\,\theta^2\w\xi^2
-L_{{3}}\,\theta^2\w\sigma^2_2
-\case16\, (3\,L_{{2}}L_{{3}}+4\,L_{{2}}
-L_{{4}}+L_{{1}}-L_{{3}}L_{{1}})\,\xi^1\w\sigma^1_2
\]\[\fl\hspace{20pt}
+\case12\,(2\,L_{{4}}-4\,L_{{2}}+1-4\,L_{{1}})\,\xi^1\w\sigma^2_2
+\case16\,(L_{{4}}-3\,L_{{2}}L_{{3}}-4\,L_{{2}}-L_{{1}}
+L_{{1}}L_{{3}})\,\xi^2\w\sigma^1_2
\]\[\fl\hspace{20pt}
+\case12\,L_{{5}}\,\xi^1\w\xi^2
+\case16\, (6\,L_{{2}}L_{{3}}+2_{{2}}+3+2\,L_{{1}}-2\,L_{{1}}L_{{3}}
-2\,L_{{4}})\,\xi^2\w\sigma^2_2,
\]
\noindent
where
\[\fl
L_{{1}}=(3\,v_x^2F_u^{2}-3\,F_{uu}v_x^{2}F-5\,v_{{x}}F_{xu}F
+5\,v_{{x}}F_{{x}}F_{{u}}-3\,u_{{x}}F_{{x}}F_{{u}}+3\,F_{uu}u_x^2F
-3\,u_x^2F_u^2
\]\[\fl\hspace{20pt}
+3\,u_{{x}}F_{xu}F)\,
(u_x^2-v_x^2)^{-1}\,F_u^{-2},
\]
\[\fl
L_{{2}}=(F_u^{2}L_1u_{{x}}+8\,v_{{x}}F_u^{2}-8\,v_{{x}}FF_{ uu}
+F_u^{2}L_1v_x)\,F_u^{-2}\,(3\,u_x-5\,v_x)^{-1},
\]
\[\fl
L_3=\case1{64}\,F_u^3\,(u_x^2-v_x^2)\,(6\,u_x^{2}F_{{u}}L_{{2}}L_{{1}}
-u_x^{2}F_{{u}}L_1^{2}
-9\,u_x^{2}F_{{u}}L_2^{2}+8\,F_{{x}}L_{{1}}v_{{x}}-24\,Fv_{{x}}L_{{2,x}}
\]\[\fl\hspace{20pt}
+8\,FL_{{1,x}}v_{{x}}
-6\,F_{{u}}v_x^{2}L_{{2}}L_{{1}}+F_{{u}}L_1^{2}v_x^{2}
-24\,F_{{x}}L_{{2}}v_{{x}}+9\,F_{{u}}v_x^{2}L_2^{2})\,v_x^{-2}
(FF_{xu}-F_xF_u)^{-2},
\]
\[\fl
L_{{4}}=-\case1{16}\,F_u\,(u_x^2-v_x^2)\, (6\,u_x^{2}F_{{u}}L_2^{2}
+9\,u_x^{2}F_{{u}}L_{{2}}-3\,u_x^{2}F_{{u}}L_{{1}}
-4\,u_x^{2}F_{{u}}L_1^{2}+10\,u_x^{2}F_{{u}}L_{{2}}L_{{1}}
\]\[\fl\hspace{20pt}
+18\,u_{{x}}F_{{x}}L_{{2}}-6\,u_{{x}}F_{{x}}L_{{1}}
-6\,F_{{u}}v_x^{2}{L_{{2}}}^{2}+10\,F_{{x}}L_{{1}}v_{{x}}
+4\,F_{{u}}L_1^{2}v_x^{2}+16\,FL_{{1,x}}v_{{x}}
\]\[\fl\hspace{20pt}
-9\,F_{{u}}v_x^{2}L_{{2}}
+3\,F_{{u}}L_{{1}}v_x^{2}-30\,F_{{x}}L_{{2}}v_{{x}}
-10\,F_{{u}}v_x^{2}L_{{2}}L
_{{1}})\,
(FF_{xu}-F_xF_u)^{-1},
\]
\noindent
while $L_5$, ..., $L_8$ are too long to be written in full. All the
differential invariants of $Cont(\Rc)$ are functions of $L_j$ and their
invariant derivatives $L_{j,I} = \Dc_I(L_j)
= \Dc_{i_1}\circ \Dc_{i_2} \circ ... \circ \Dc_{i_l}(L_j)$
(the operators $\Dc_i$ are not the same as in the case $\Pc_1$!) The $s^{th}$
order classifying manifold associated with the coframe $\bftheta$ and an open
subset $V$ is
\begin{equation}\fl
\Cc^{(s)}(\bftheta,V) = \{(L_{j,I}(x,u,u_x,v_x))\,\,\vert\,\,
j\in\{1,...,8\},\,\,
\#I\le s, \,\,(x,u,u_x,v_x)\in V\}.
\label{manifold_B1}
\end{equation}
\noindent Since all the functions $L_{j,I}$ depend on four
variables $x$, $u$, $u_x$, and $v_x$, it follows that $\rho_s =
\dim \Cc^{(s)}(\bftheta,V) \le 4$ for all $s\ge 0$, and the order
$r$ of the coframe $\bftheta$ satisfies $r+1 \le 4$ again. Two
systems (\ref{B}) from the subclass $\Pc_2$ are locally equivalent
under a contact transformation if and only if their fourth order
classifying manifolds (\ref{manifold_B1}) locally overlap, and
$\dim Cont(\Rc)=6 - \dim \Cc^{(4)}(\bftheta,V)\ge 2$.

\vskip 5 pt

If $(\ln F)_{xu} = 0$, then $F(x,u) = S(x)\,\tilde{F}(u)$ for arbitrary
functions $S$ and $\tilde{F}$. Then the change of variables $\tilde{x} = H(x)$
provided $H^{\prime}(x) = (S(x))^{-1}$ maps the system
$u_t = S(x)\,\tilde{F}(u)\,v_x$,
$v_t = S(x)\,\tilde{F}(u)\,u_x$,
to the system
$u_t = \tilde{F}(u)\,v_{\tilde{x}}$,
$v_t = \tilde{F}(u)\,u_{\tilde{x}}$. We drop the tildes for simplicity of
notation and consider the system
\begin{equation}
u_t = F(u)\,v_x,\qquad
v_t = F(u)\,u_x.
\label{F}
\end{equation}
\noindent
The computations show that there are three non-equivalent types of systems
(\ref{F}): denote by $\Pc_3$ the subclass of systems (\ref{F}) such that
$F_u \not = 0$ and
\[
M_1={{4FF_u^2F_{uu}+4F^2F_{uu}^2 -4F^2F_uF_{uuu}-3F_u^4}\over{F_u^4}}
\not= const,
\]
by $\Pc_4$ denote the subclass of systems (\ref{F}) such that
$F_u \not = 0$ and $M_1=const$, finally,  by $\Pc_5$ denote the subclass
of systems (\ref{F}) such that $F_u=0$.

\vskip 5 pt

The subclass $\Pc_3$ is not empty; for example, system (\ref{F}) with
$F(u)= (1+u^2)^{-1}$ belongs to $\Pc_3$. For a system from $\Pc_3$ the
structure equations of the symmetry pseudo-group after a prolongation have the
form
\[\fl
d\theta^1=
\eta_1\w\theta^1
-\theta^2\w\xi^1
+\xi^2\w\sigma^1_2,
\]
\[\fl
d\theta^2=
\eta_1\w\theta^2
+\theta^1\w\xi^2
-M_2\,\theta^2\w\xi^1
-(2\,M_2+M_3)\,\theta^2\w\xi^2
+(\xi^1+\xi^2)\w\sigma^2_2,
\]
\[\fl
d\xi^1=
M_3\,\xi^1\w\xi^2,
\]
\[\fl
d\xi^2=
M_2\,\xi^1\w\xi^2,
\]
\[\fl
d\sigma^1_2=
\eta_1\w\sigma^1_2
+\eta_2\w\xi^2
+M_1\,\theta^1\w\xi^1
-2\,M_2\,\theta^2\w\xi^1
-M_2\,\xi^1\w\sigma^1_2
+\xi^1\w\sigma^2_2,
\]
\[\fl
d\sigma^2_2=
\eta_1\w\sigma^2_2
+\eta_3\w(\xi^1+\xi^2)
+M_1\,\theta^2\w\xi^2
-2\,(M_2+M_3)\,\xi^2\w\sigma^2_2,
\]
\[\fl
d\eta_1=
(M_1-1)\,\xi^1\w\xi^2,
\]
\[\fl
d\eta_2=
\mu_1\w\xi^2
+\eta_1\w\eta_2
+2\,M_2\,\eta_2 \w\xi^1-\eta_3\w\xi^1
+\left(\Dc_4(M_1)+2M_2-M_1M_3\right)\,\theta^1\w\xi^1
\]\[\fl\hspace{20pt}
-\left(2\Dc_4(M_2)+4M_2^2-M_1\right)\,\theta^2\w\xi^1
+\left(\Dc_4(M_2)+2M_1-M_2M_3-1\right)\,\xi^1\w\sigma^1_2
\]\[\fl\hspace{20pt}
-(4M_2+M_3)\,\xi^1\w\sigma^2_2,
\]
\[\fl
d\eta_3=
\mu_2\w(\xi^1+\xi^2)
+\eta_1\w\eta_3
-3\,(M_2+M_3)\,\eta_3\w\xi^2
-(2M_1M_2+1)\,\theta^2\w\xi^2
\]\[\fl\hspace{20pt}
+\left(4M_1-2\Dc_3(M_2)(\Dc_4(M_1)-1)+2M_2(M_2+M_3)-3\right)\,
\xi^2\w\sigma^2_2,
\]
\noindent
where $\eta_1$, $\eta_2$, $\eta_3$, $\mu_1$, and $\mu_2$ are 1-forms on
$\Rc\times\Hc$. The only non-zero {\it reduced character},
\cite[def~11.4]{Olver95}, is $s_1^{\prime} =2$, therefore the symmetry
pseudo-group for system (\ref{F}) from $\Pc_3$ depends on two arbitrary
functions of one variable. The invariants $M_2$ and $M_3$ are defined by
$M_2 =(2FF_{uu}M_{1,u}-FF_uM_{1,uu}-2F_u^2M_{1,u})F^{-1}F_u^{-1}M_{1,u}^{-2}$
and $M_3 = -(M_2\,\Dc_4(M_1)+\Dc_{(3,4)}(M_1))$, where for an arbitrary
function $R(u)$ we have $dR =\Dc_3(R)\,\xi^1+\Dc_4(R)\,\xi^2$ with the
invariant derivatives $\Dc_3=M_{1,u}^{-1}\,\partial/\partial u$ and
$\Dc_4=(1-4F^2M_{1,u}^2F_u^{-2})\,M_{1,u}^{-1}\,\partial/\partial u$.
We have $\Dc_3(M_1)=1$ and $\Dc_4(M_1) = 1-4F^2M_{1,u}^2F_u^{-2}$. Since
$M_1\not = const$, then $M_2$ and $\Dc_4(M_1)$ depend on $M_1$ functionally:
$M_2 = H_1(M_1)$ and $\Dc_4(M_1) = H_2(M_1)$. All the other differential
invariants can be expressed as functions of $M_1$. For example, we have
$\Dc_3(M_2) = H_1^{\prime}(M_1)\,\Dc_3(M_1) = H_1^{\prime}(M_1)$ and
$\Dc_4(M_2) = \Dc_4(M_1)\,\Dc_3(M_2) = H_2(M_1)\,H_1^{\prime}(M_1)$.

The first order classifying manifold associated with the coframe
$\bftheta=\{\theta^1, \theta^2, \xi^1, \xi^2, \sigma^1_2,
\sigma^2_2, \eta_1, \eta_2, \eta_3\}$ and an open subset $W
\subset \R$ can be parameterized by $M_1$, $M_2$, and $\Dc_4(M_1)$
:
\begin{equation}
\Cc^{(1)}(\bftheta,W) = \{(M_1(u),M_2(u),\Dc_4(M_1)(u))\,\,\vert\,\,
 \,\,u\in W\}.
\label{manifold_B2}
\end{equation}
\noindent
Two systems (\ref{F}) from $\Pc_3$ are equivalent under a contact
transformation iff their classifying manifolds (\ref{manifold_B2})
(locally) overlap, \cite[Th~15.22]{Olver95}, i.e., they have the same
functions $H_1$ and $H_2$.

\vskip 5 pt

The subclass $\Pc_4$ is not empty; for example, systems (\ref{F})
with $F(u) = \exp(C\arctan(\sinh(\lambda\,u)))$, $F(u) = \e^u$, or
$F(u) = u^m$, $m\not= 0$, belong to $\Pc_4$. For a system from
$\Pc_4$ the structure equations of symmetry pseudo-group after a
prolongation have the form
\[\fl
d\theta^1=
\eta_1\w\theta^1
-\theta^2\w\xi^1
+\xi^2\w\sigma^1_2,
\]
\[\fl
d\theta^2=
\eta_2\w\theta^2
-\theta^1\w\xi^2
+(\xi^1+\xi^2)\w\sigma^2_2,
\]
\[\fl
d\xi^1=
(\eta_1-\eta_2)\w(\xi^1+2\,\xi^2),
\]
\[\fl
d\xi^2=
(\eta_2-\eta_1)\w\xi^2,
\]
\[\fl
d\sigma^1_2=
-2\,\eta_1\w(\theta^2-\sigma^1_2)
+\eta_2\w (2\,\theta^2-\sigma^1_2)
+\eta_3\w\xi^2
+M_1\,\theta^1\w\xi^1
+\xi^1\w\sigma^2_2,
\]
\[\fl
d\sigma^2_2=
(2\,\eta_2-\eta_1)\w\sigma^2_2
+\eta_4\w(\xi^1+\xi^2)
+M_1\,\theta^2\w\xi^2,
\]
\[\fl
d\eta_1=
(M_1-1)\,\xi^1\w\xi^2,
\]
\[\fl
d\eta_2=
-(M_1-1)\,\xi^1\w\xi^2,
\]
\[\fl
d\eta_3=
\mu_1\w\xi^2
+(3\,\eta_1-2\,\eta_2)\w\eta_3
-2\,(M_1+1)\,(\eta_1-\eta_2)\w\theta^1
+4\,(\eta_1+\eta_2)\w\sigma^2_2
-\eta_4\w\xi^1
\]\[\fl\hspace{20pt}
+(3\,M_1-4)\,\theta^2\w\xi^1
+(4\,M_1-3)\,\xi^1\w\sigma^1_2,
\]
\[\fl
d\eta_4=
\mu_2\w(\xi^1+\xi^2)
-(2\,\eta_1-3\,\eta_2)\w\eta_4
+ (4\,M_1-3)\,\xi^2\w\sigma^2_2,
\]
\noindent
where $\eta_1$, $\eta_2$, $\eta_3$, $\eta_4$, $\mu_1$, and $\mu_2$ are 1-forms
on $\Rc\times\Hc$. The only non-zero reduced character is $s_1^{\prime} =2$,
therefore the symmetry pseudo-group for system (\ref{F}) from $\Pc_3$ depends
on two arbitrary functions of one variable. Since $M_1=const$, all the other
differential invariants are equal to zero, and the classifying manifold is
a point. Two systems from $\Pc_4$ are equivalent under a contact transformation
iff they have the same values of $M_1$.

\vskip 5 pt

A system from $\Pc_5$ with $F(u)\equiv m = const$ can be
transformed to the system
\begin{equation}
u_t = v_x,\qquad v_t = u_x
\label{linear_wave}
\end{equation}
\noindent
by the change of variables $t\mapsto m^{-1}\,t$. The symmetry pseudo-group
for system (\ref{linear_wave}) has the structure equations
\[
d\theta^1=\eta_1\w\theta^1+\xi^1\w\sigma^1_2,
\]
\[
d\theta^2=\eta_2\w\theta^2+\xi^2\w\sigma^2_2,
\]
\[
d\xi^1=
\eta_3\w\xi^1
+\eta_4\w\theta^1,
\]
\[
d\xi^2=
\eta_5\w\xi^2
+\eta_6\w\theta^2,
\]
\[
d\sigma^1_2=
(\eta_1-\eta_3)\w\sigma^1_2
+\eta_7\w\theta^1
+\eta_8\w\xi^1,
\]
\[
d\sigma^2_2=
(\eta_2
-\eta_5)\w\sigma^2_2
+\eta_9\w\theta^2
+\eta_{10}\w\xi^2,
\]
\noindent
where $\eta_1$, ..., $\eta_{10}$ are 1-forms on $\Rc\times\Hc$. The non-zero
reduced characters are $s_1^{\prime}= 6$ and $s_2^{\prime}= 4$, therefore
the pseudo-group depends on 4 arbitrary functions of two variables.

The subclasses $\Pc_3$ and $\Pc_4$ are linearizable: the contact
transform $\tilde{t}=v$, $\tilde{x}=u$, $\tilde{u}=x$, and
$\tilde{v}=t$ maps system (\ref{F}) to the system
$\tilde{u}_{\tilde{t}} = F(\tilde{x})\,\tilde{v}_{\tilde{x}}$,
$\tilde{v}_{\tilde{t}} =
(F(\tilde{x}))^{-1}\,\tilde{u}_{\tilde{x}}$. Therefore all systems
(\ref{Main}) with infinite-dimensional symmetry pseudo-groups are
linearizable, cf. \cite{ReidWittkopf}.

\vskip 5 pt
The results of the computations are summarized in the following

\vskip 5 pt
\noindent{\bf Theorem :}
{\it Every system from the class of nonlinear wave equations (\ref{Main}) is
equivalent under a contact transformation to a system from one of the
five invariant subclasses} $\Pc_1$, $\Pc_2$, $\Pc_3$, $\Pc_4$, and  $\Pc_5$:
$\Pc_1$ {\it consists of all systems (\ref{Main}) such that} $G_x\not=0$,
$F_u\not=0$, {\it and} $G_x\,F^2\,F_u^{-1}\not=const$,
$\Pc_2$ {\it consists of all systems (\ref{B}) such that}
$(\ln F)_{xu}\not = 0$,
$\Pc_3$ {\it consists of all systems (\ref{F}) such that}
$M_1=(4FF_u^2F_{uu}+4F^2F_{uu}^2 -4F^2F_uF_{uuu}-3F_u^4)\,F_u^{-4}\not= const$,
$\Pc_4$ {\it consists of all systems (\ref{F}) such that} $M_1 = const$,
and $\Pc_5$ {\it consists of system (\ref{linear_wave})}.

{\it Systems from} $\Pc_1$ {\it and} $\Pc_2$ {\it have
finite-dimensional symmetry groups, while systems from} $\Pc_3$,
$\Pc_4$, {\it and} $\Pc_5$ {\it are linearizable and have
infinite-dimensional symmetry pseudo-groups.}

{\it Two systems from one of the subclasses} $\Pc_1$, $\Pc_2$, {\it or} $\Pc_3$
{\it are equivalent to each other under a contact transformation if and only if
the classifying manifolds (\ref{manifold_A}), (\ref{manifold_B1}), or
(\ref{manifold_B2}) for these systems locally overlap. Two systems from
the subclass} $\Pc_4$ {\it are equivalent if and only if they have the same
constant value of the invariant} $M_1$.

\section*{Conclusion}
In this paper, the moving coframe method of \cite{FO} is applied
to the local equivalence problem for a class of systems of
nonlinear wave equations under an action of the pseudo-group of
contact transformations. We have found five invariant subclasses
and shown that every system of nonlinear wave equations can be
transformed to a system from one of these subclasses. The
structure equations and the differential invariants for all the
subclasses are found. The solution of the equivalence problem is
given in terms of the differential invariants. Three of the
invariant subclasses consist of linearizable systems with
infinite-dimensional symmetry pseudo-groups. Therefore all the
linearizable cases for non-linear wave equations are classified.

\end{document}